\documentclass[reprint,prl,aps,superscriptaddress]{revtex4-1}
\usepackage{times}
\usepackage{graphicx}
\usepackage{amsmath, braket, amsfonts}
\usepackage{amssymb}
\usepackage{natbib}
\usepackage{sidecap}
\usepackage{bm, color, ulem}
\usepackage{xcolor}
\usepackage{ragged2e}
\usepackage{wrapfig,lipsum,booktabs}
\usepackage{siunitx}

\newcommand\dif{\mathop{}\!\mathrm{d}}
\newcommand{\be}{\begin{equation}}
\newcommand{\ee}{\end{equation}}

\DeclareUnicodeCharacter{03BD}{{$\nu$}}
\makeatletter
\def\maketitle{
\@author@finish
\title@column\titleblock@produce
\suppressfloats[t]}
\makeatother

\begin{document}
\title{Spontaneous localization at a potential saddle point \\ from edge state reconstruction in a quantum Hall point contact}

\author{Liam A. Cohen}
\thanks{These authors contributed equally to this work}
\affiliation{Department of Physics, University of California at Santa Barbara, Santa Barbara CA 93106, USA}
\author{Noah L. Samuelson}
\thanks{These authors contributed equally to this work}
\affiliation{Department of Physics, University of California at Santa Barbara, Santa Barbara CA 93106, USA}

\author{Taige Wang}
\affiliation{Department of Physics, University of California, Berkeley, California 94720, USA}
\affiliation{Material Science Division, Lawrence Berkeley National Laboratory, Berkeley, California 94720, USA}
\author{Kai Klocke}
\affiliation{Department of Physics, University of California, Berkeley, California 94720, USA}
\author{Cian C. Reeves}
\affiliation{Department of Physics, University of California at Santa Barbara, Santa Barbara CA 93106, USA}
\author{Takashi Taniguchi}
\affiliation{International Center for Materials Nanoarchitectonics,
National Institute for Materials Science,  1-1 Namiki, Tsukuba 305-0044, Japan}
\author{Kenji Watanabe}
\affiliation{Research Center for Functional Materials,
National Institute for Materials Science, 1-1 Namiki, Tsukuba 305-0044, Japan}
\author{Sagar Vijay}
\affiliation{Department of Physics, University of California at Santa Barbara, Santa Barbara CA 93106, USA}
\author{Michael P. Zaletel}
\affiliation{Department of Physics, University of California, Berkeley, California 94720, USA}
\affiliation{Material Science Division, Lawrence Berkeley National Laboratory, Berkeley, California 94720, USA}
\author{Andrea F. Young}
\email{andrea@physics.ucsb.edu}
\affiliation{Department of Physics, University of California at Santa Barbara, Santa Barbara CA 93106, USA}

\date{\today}

\begin{abstract}
Quantum point contacts (QPCs) are an essential component in mesoscopic devices. Here, we study the transmission of quantum Hall edge modes through a gate-defined QPC in monolayer graphene. We observe resonant tunneling peaks and a nonlinear conductance pattern characteristic of Coulomb-blockaded localized states. 
The in-plane electric polarizability reveals the states are localized at a classically-unstable electrostatic saddle point. We explain this unexpected finding within a self-consistent Thomas-Fermi model, finding that localization of a zero-dimensional state at the saddle point is favored whenever the applied confinement potential is sufficiently soft compared to the Coulomb energy. Our results provide a direct demonstration of Coulomb-driven reconstruction at the boundary of a quantum Hall system.
\end{abstract}
\maketitle

Quantum point contacts (QPCs) operated in the quantum Hall regime ubiquitously show resonant tunneling features as edge modes are successively pinched off. These features are generally assumed to arise from uncontrolled disorder potentials introduced, for example, during the fabrication of local split gates\cite{baer_interplay_2014, milliken_indications_1996, deprez_tunable_2021, zimmermann_tunable_2017, ronen_aharonov-bohm_2021}. 
Here we present evidence that repeatable resonant tunneling features can also arise independently of local disorder via the same Coulomb-interaction driven mechanism that leads to edge-state reconstruction\cite{chamon_sharp_1994}.  
Theoretically, electrostatic confinement in the  quantum Hall regime is characterized by the confinement energy, $E_V = e\frac{d\phi_{ext}}{dx}\ell_B$, where $\phi_{ext}$ is the applied electric potential. 
 The spatial structure of the electron density in the presence of an external potential is controlled by the dimensionless ratio between $E_V$ and the Coulomb energy, $E_C = \frac{e^2}{4\pi \epsilon \ell_B}$. In the sharp-edge regime, $E_V/E_C >> 1$, the electron density is expected to decrease sharply and monotonically at the edge of the quantum Hall bulk, leading to the simplest possible (unreconstructed) edge structure. At lower sharpness, corresponding to $E_V/E_C \sim 1$, a delicate balance between the Coulomb interaction and confining electric potential leads to a more complicated spatial structure in the electron density, which may include a nonmonotonic dependence of the density on position.  
 
Edge-state reconstruction is often invoked as a complicating factor in experiments seeking to probe universal properties of fractional quantum Hall states. 
For example, the emergence of charge-neutral modes related to edge reconstruction is tied to the suppression of visibility in edge-state interference experiments\cite{bhattacharyya_melting_2019} or the unexpected observation of upstream heat-transport in certain experiments\cite{sabo_edge_2017,venkatachalam_local_2012}.  
Universal, reconstruction-free edges\cite{hu_realizing_2011} have been observed using scanning tunneling microscopy at the physical edge of graphene systems\cite{li_evolution_2013,coissard_absence_2023}, where the confining potential is atomically sharp.  
However direct measurements of the edge profile have been elusive at the softer, gate-defined edges that constitute the essential ingredient for mesoscopic devices, such as edge state interferometers, that allow for \textit{in situ}  edge state control.

Here, we study a monolayer graphene device fitted with an electrostatically tunable quantum point contact. 
Our geometry allows the sharpness of the confinement potential at the QPC to be controlled \textit{in situ}, allowing us to directly probe the nature of the edge through its effects on the transmission of charge through the point contact. 
Our main result is an unexpected manifestation of Coulomb-induced reconstruction: the spontaneous formation of a quantum dot localized at an electrostatic saddle point. 
Our observations imply a local discrepancy between the \textit{applied} electrostatic potential set by the voltages applied to each gate electrode and the \textit{total} electronic potential, which takes into account electron-electron interactions and the finite compressibility of the 2DEG at high magnetic fields.

\begin{figure*}[ht]
    \centering
    \includegraphics[width = 180mm]{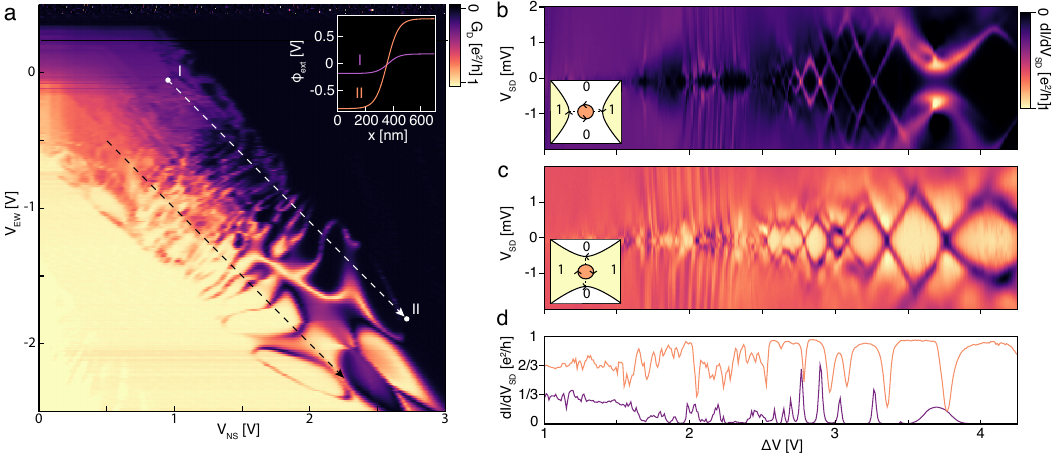}
    \caption{\textbf{Coulomb blockade at a gate-induced saddle point.} 
    \textbf{(a)} $G_D$ at $V_B = -0.25$~V and B=13~T spanning zero to full transmission of a single IQH edge mode. 
    Inset: applied potential $\phi_{ext}$, as determined by electrostatic simulations, across the N-W boundary at the points marked I and II in the main panel. 
    $E_V/E_C\approx 0.4, 1.8$ for I and II respectively. 
    \textbf{(b)} Two-terminal differential conductance $\dif I/ \dif V_{SD}$ measured across the QPC. Data are plotted as a function of $V_{SD}$ along the white contour in panel a, parameterized by $\Delta V = V_{NS} - V_{EW}$.
    Diamond-shaped conductance peaks are indicative of Coulomb blockade, consistent with resonant transmission of edge modes through a localized state at the center of the QPC, as depicted in the inset. 
    \textbf{(c)} $\dif I/ \dif V_{SD}$ along the black contour in panel a.  Resonant reflection is observed, consistent with backscattering of edge state through the localized state, depicted in the inset. 
    \textbf{(d)} Zero-bias traces of (b) and (c) illustrate that the resonant transmission and resonant reflection regimes evolve from fractional plateaus at 1/3 and 2/3 respectively as $\Delta V = V_{NS} - V_{EW}$ is increased.}
    \label{fig:dot_antidot}
\end{figure*}

We study a  dual-graphite gated hBN-encapsulated graphene van der Waals heterostructure described previously (see \cite{cohen_nanoscale_2023,cohen_universal_2023} and supplementary information). Local anodic oxidation lithography \cite{li_electrode-free_2018} is used to pattern an ``X'' shape in a graphite flake, which is then integrated into the van der Waals heterostructure\cite{cohen_nanoscale_2023}.   
A subsequent reactive ion etch defines the overall device outline, electrically separating the graphite into four separate gates which we label north (N), south (S), east (E) and west (W).  
Together, these tune the electron density in each of four quadrants separately. The E and W gates are used to tune the electron density on either side of the device while the N and S gates deplete the electron density to form the QPC constriction.  The graphite bottom gate, controlled by a voltage $V_{B}$, can be used in conjunction with the top gates to control the fringe-fields at the boundaries between all four regions while maintaining a fixed electron density within the bulk of each quadrant, directly tuning $E_V/E_C$ along the edge and at the potential saddle point. This fabrication method, by avoiding subtractive processing in the critical regions of the completed device, results in significantly lower disorder within the quantum point contact\cite{cohen_nanoscale_2023, cohen_universal_2023}. 

To characterize the operation of our QPC we excite an AC voltage on one side of the device and measure the current $I$ into a contact on the opposite side of the QPC and the diagonal voltage drop $V_D$ to calculate the four terminal ``diagonal conductance'', $G_D \equiv I/V_D$ (see supplementary information). In the integer quantum Hall (IQH) regime this quantity gives a direct measure of the number of edge modes $N_{qpc}$ transmitted across the device such that $G_D = N_{qpc} e^2/h$\cite{datta_electronic_1995, zimmermann_tunable_2017}.

Fig.~\ref{fig:dot_antidot}a shows $G_D$ measured over a large range of $V_N = V_S = V_{NS}$ and $V_E = V_W = V_{EW}$ at B=13T and $T = \SI{300}{mK}$ with $\nu_{EW} \leq 0$.  The plot is centered around the transition between $G_D=0$ and $G_D = e^2/h$. The lower left of the plot corresponds to full transmission of the outermost integer quantum Hall edge mode, while lower right corresponds to full pinch-off.  The bulk filling factor varies from $\nu_{EW} = 0$ at $V_{EW} = 0.5$ to $\nu_{EW} = -2$ at $V_{EW} = -2.5V$.  Despite the range of bulk filling factors spanned in Fig.~\ref{fig:dot_antidot}, $G_D$ varies only from $0$ to $e^2/h$ indicating that only the outermost integer quantum Hall edge mode is transmitted across the device. 

Tuning the gate voltages from upper left to lower right in Fig. 1a, corresponding to lines of constant $V_{NS} + V_{EW}$, leaves the applied potential in the middle of the QPC constant.  Naively, we expect $G_D$ to also remain constant; however, we observe intricate structure to the measured conductance characterized by the emergence of sharp conductance peaks and dips at the plateau transition.  These features are tuned both by the sum and difference of $V_{NS}$ and $V_{EW}$.  

We attribute this structure to the reconstruction of the saddle point potential due to electron-electron  interactions. The inset to Fig.~\ref{fig:dot_antidot}a shows finite-element simulations of the applied electrostatic potential across the boundary between N and W regions, with the gate voltages set to the values given at the points labelled I and II in Fig. 1a.  Within our  simulations, $E_V/E_C$ can be tuned by more than a factor of four between points I and II, significantly changing the sharpness of the confinement.  It is natural to attribute the resonant structure to the effect of this variable; evidently, transmission across the device is highly sensitive to the  sharpness of the potential at the  the saddle point.

A notable feature of the data in Fig. 1a is the approximate reflection symmetry across the expected line of $G_D=\frac{1}{2}e^2/h$ that relates the two marked trajectories.  
Figs. \ref{fig:dot_antidot}b and c show the differential conductance across the quantum point contact, $\dif I/ \dif V_{SD}$, plotted as a function of the source-drain voltage $V_{SD}$ as well as the coordinate $\Delta V = V_{NS} - V_{EW}$ parameterizing the contours shown in Fig.~ \ref{fig:dot_antidot}a. Along the white contour the junction is nearly pinched off, with conductance dropping to zero between high conductance peaks.  Nonlinear conductance in this regime shows diamond structure typical of transport across a Coulomb blockaded quantum dot, with charging energies as large as 1 meV.  This is consistent with a scenario where resonant transmission through a dot, located at the QPC center, allows charge transport between two otherwise fully-reflected edge states in the E and W regions.  This scenario is illustrated schematically in the inset to Fig.~\ref{fig:dot_antidot}b. 

Along the dashed black arrow in Fig.~\ref{fig:dot_antidot}a, nonlinear conductance shows an almost identical diamond structure, except this time with the on-resonance condition corresponding to a decrease in conductance (Fig.~\ref{fig:dot_antidot}c). This is again consistent with a Coulomb blockaded quantum dot in the QPC, but one whose effect on transport is to allow backscattering between two otherwise fully-transmitted edge states (see Fig.~\ref{fig:dot_antidot}c, inset). 

In both scenarios, as $\Delta V$ is increased, or equivalently, as the potential sharpness is increased, the level spacing of the quantum dot increases.  This is consistent with the size of the quantum dot decreasing with increasing $\Delta V$, either making the quantum mechanical level spacing smaller, or simply decreasing the island capacitance leading to an increase in the charging energy.  Additionally, in Figs.~\ref{fig:dot_antidot}b and c, the stability of the Coulomb diamonds decreases with $\Delta V$, eventually giving rise to fractional plateaus at $G_D = \frac{1}{3} e^2/h$ and $G_D = \frac{2}{3} e^2/h$.  This is highlighted in Fig.~\ref{fig:dot_antidot}d which plots zero-bias data extracted from Figs.~\ref{fig:dot_antidot}b and c.

\begin{figure}
    \includegraphics[width = 90mm]{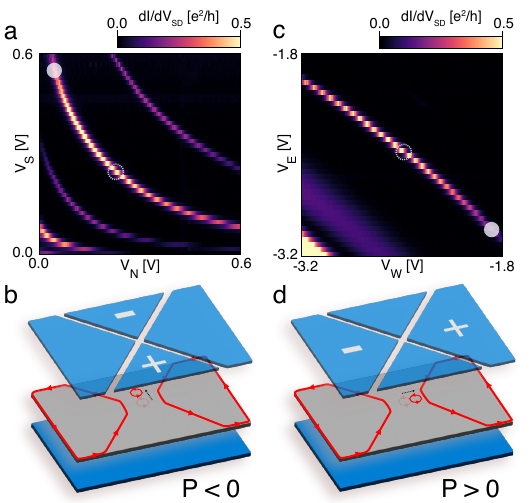}
    \caption{\textbf{Quantum dot position and polarizability.} 
    \textbf{(a)} $\dif I/\dif V_{SD}$ at $V_{SD}=0$ across several Coulomb blockade peaks, plotted as a function of $V_N$ and $V_S$ with other voltages held constant.  Near $V_N=V_S$, $\dif V_S/\dif V_N\approx-1$ indicating that the dot is equidistant from the N and S gates.  The peak follows a contour of positive curvature, indicating the dot is repelled by positive $V_N$ or $V_S$. \textbf{(b)}  Cartoon schematic demonstrating a negative quantum dot polarizability as measured in panel a.  Here the trapped charge moves in the opposite direction as the applied electric dipole moment.
    \textbf{(c)} The same, but plotted in the $(V_W,V_E)$ plane. Again, $\dif V_E/\dif V_W\approx-1$ indicates that the dot is equidistant from the E and W gates at zero detuning.  The curvature is negative, opposite to that in the $(V_N,V_S)$ plane. \textbf{(d)} Cartoon schematic demonstrating a positive quantum dot polarizability as measured in panel c.  Here the trapped charge moves along the applied electric dipole moment.  The opposite sign of polarizability along orthogonal directions for the same Coulomb blockade peak indicates the localized charge tracks the equipotentials of the gate-defined saddle point.}
    \label{fig:dot_polarizability}
\end{figure}

The existence of a quantum dot is not na\"ively expected in a quantum point contact where the unscreened electrostatic potential realizes a saddle point.  The question thus arises as to whether such behavior can arise intrinsically, or if electron confinement might arise from the presence of uncontrolled impurity potentials near the quantum point contact.  
To address this question, we  use the four-quadrant gate geometry \cite{cohen_nanoscale_2023} to determine both the position of the localized charge and its polarizability within the 2D plane.  
Fig.~\ref{fig:dot_polarizability}a shows several representative Coulomb blockade peaks as function of $V_N$ and $V_S$ with all other gate voltages constant. The Coulomb peaks follow the condition $C_N  \mathrm{d} V_N + C_S \mathrm{d} V_S  = 0$, where $C_N$ and $C_S$ are the capacitances to the N and S gates, respectively. 
The ratio of these capacitances may be inferred from the slope of the Coulomb peak trajectory in the $(V_N,V_S)$ plane,  $\frac{ \mathrm{d} V_S}{ \mathrm{d} V_N} = - \frac{C_N}{C_S}$. 
When $V_N = V_S$, we find that $C_N/C_S\approx1$, implying that the dot is equidistant from the two gates. 
The curvature of the peak trajectory, meanwhile, reveals how the capacitances are changed by the motion of the dot.  
Taking $C_N = C_S$, the curvature is defined by the expression  $\frac{ \mathrm{d}^2 V_S}{ \mathrm{d} V^2_N} = - \frac{1}{C_S} ( \frac{\dif C_N}{\dif V_N} - \frac{\dif C_S}{\dif V_N})$. In Fig. 2a, the curvature is observed to be positive, implying $\frac{\dif C_N}{\dif V_N} < \frac{\dif C_S}{\dif V_N}$. Since $C_{N/S}$ is inversely proportional to the distance between the gate and dot, this implies that a positive bias on $V_{N}$ repels the dot from the $N$ gate, as schematically illustrated in Fig.~\ref{fig:dot_polarizability}b. This behavior is consistent with a charge trapped at an electrostatic \textit{minimum} in the N/S direction. 

Analogous measurements as a function of $V_E$ and $V_W$ are shown in Fig.~\ref{fig:dot_polarizability}c.  Again, for $V_E = V_W$, $C_E/C_W\approx 1$, indicating the dot is equidistant from the E and W gates.  However, the curvature of the peak trajectories in Fig.~\ref{fig:dot_polarizability}c are opposite in sign from those in Fig.~\ref{fig:dot_polarizability}a. This behavior is consistent with a particle trapped in an electrostatic \textit{maximum} along the E/W direction, as illustrated in Fig.~\ref{fig:dot_polarizability}d. 
Taken together, these measurements show that the quantum dot is centered at an electrostatic saddle point.  This behavior is consistent across different sets of resonances, different gate voltages, and different magnetic fields, as shown in the Supplementary information. 
The same characteristic behavior is observed for partial transmission of a quantum Hall edge mode in the \textit{absence} of resonant behavior, again described in the supplementary.  The similarity between the behavior of the quantum dot and the delocalized edge mode supports the idea that the charge in the dot localized precisely at the saddle point.  This is a surprising result given that a saddle point potential does not support localized states even in a magnetic field \cite{hegde_quasinormal_2019, floser_transmission_2010}. 

These obsevations can be understood by considering the effects of the Coulomb interaction.  Generally, edge state reconstruction may lead to non-monotonic density profiles along a smooth potential step \cite{chamon_sharp_1994,khanna_fractional_2021}. 
Along the translation-invariant electrostatic edge between two quantum Hall phases, the resulting electronic density retains the spatial symmetry of the underlying potential, resulting in formation of a series of strips at the boundary between the two phases. 
In a more complex geometry such as the QPC potential studied here, the same mechanism can favor the formation of more complicated structures that still maintain the 180$^\circ$-rotation symmetry of the underlying potential, such as an isolated dot of nonzero density at the center of the QPC.

\begin{figure}[ht]
    \centering
    \includegraphics[width = 90mm]{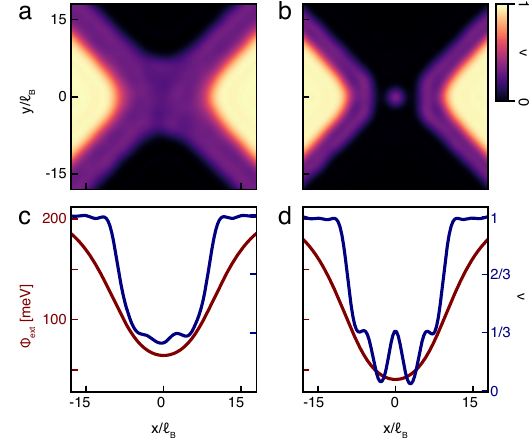}
    \caption{
    \textbf{Thomas-Fermi simulations.} 
    \textbf{(a)} $\nu$ calculated for $B = 13T$ and $E_V/E_C = 0.44$.  The continuous strip of $\nu=\tfrac13$ corresponds to transmission of a single e/3 mode, as is observed at point I in Fig.~\ref{fig:dot_antidot}a.  
    \textbf{(b)} $\nu$ calculated for $E_V/E_C = 0.49$.  
    For this range of parameters, a small island of $\nu=\tfrac13$ forms at the center of the QPC, separated by depletion regions with $\nu=0$ from the E and W regions consistent with the Coulomb blockade observed at point II in Fig.~\ref{fig:dot_antidot}.
     \textbf{(c)} The unscreened potential $\Phi_{ext}$ induced externally by the applied gate voltages compared with the reconstructed carrier density $\nu$ as a function of $x$ with $y=0$ for $E_V/E_C = 0.44$ 
    \textbf{(d)} $\Phi_\textrm{ext}$ and $\nu$ for $E_V/E_C = 0.49$. 
    }
    \label{fig:TF_reconstruction}
\end{figure}

To evaluate the plausibility of a reconstruction-induced quantum dot we use a self-consistent Thomas-Fermi model to calculate the reconstructed density within the QPC.  Our  calculations account for the density-dependent chemical potential within a partially-filled LL extracted from thermodynamic experiments\cite{yang_experimental_2021}, as well as a realistic device geometry (see supplementary information). 
Fig.~\ref{fig:TF_reconstruction}a-b shows the calculated density for $\nu_{EW} = 1$ and $\nu_{NS} = 0$. 
For the softest potential, in panel a, an intermediate side-strip of fractional filling $\nu = 1/3$ is observed at each boundary between $\nu = 0$ and $\nu = 1$. 
As $E_V/E_C$  increases,  reconstruction  becomes less favorable, and an isolated island of filling factor $\nu=1/3$ spontaneously forms in the center of the junction. Crucially, this island is isolated from the surrounding regions of non-zero density by depletion regions with $\nu\approx 0$.  
In a transport experiment, these regions may form tunnel barriers between the dot and the reservoirs on either side. 

Both regimes of reconstruction correspond directly to observations in Fig.~\ref{fig:dot_antidot}.  At point I along the dashed white line, where $\Delta V = V_{NS} - V_{EW} = 1 V$ and the applied potential is thus the softest, we observe fractionally-quantized conductance, $G_D = 1/3$, corresponding to transmission of a single fractional edge mode, in agreement with Fig.~\ref{fig:TF_reconstruction}a. Additional evidence for the existence of incompressible strips at fractional filling factors within the QPC, maintained over a wide array of electrostatic configurations, is given in supplementary Fig.~5.  As $\Delta V$ is increased to $Delta V \sim 2.5V$, Coulomb peaks appear corresponding to the existence of a quantum dot as in Fig.~\ref{fig:TF_reconstruction}b. Simulations show that the size of the quantum dot shrinks with further increase of $\Delta V$, while the reflected edge mode still maintains strips of $\nu = 1/3$ away from the QPC, leading to an increase in the level spacing as observed in Fig.~\ref{fig:dot_antidot}b-d.  At sufficiently large $\Delta V$, reconstruction should no longer be favorable anywhere within the device, and one is expected to recover the sharp-edge limit with a single chiral edge mode for a bulk $\nu = 1$ quantum Hall state. As described in the supplementary information, data from a second, identical device shows resonant structure for $E_V / E_C < 2.2$ but a single monotonic step in $G_D$ for $E_V / E_C > 2.2$. Evidently, for appropriate electrostatic conditions the non-reconstructed limit can be recovered in experiment even for gate defined edges.

Interestingly, within our simulations the quantum dot is composed of an island at fractional filling.  This suggests that single fractionally charged quasiparticles can be localized even in a geometry which forbids trapping single particles in the absence of Coulomb interactions.  Additionally, in light of our simulations the highly symmetric nature of the resonant reflected features observed in Fig.~\ref{fig:dot_antidot}a may likely be interpreted as the particle-hole conjugate to the scenario presented in Fig.~\ref{fig:TF_reconstruction}b. Future experiments, for example measuring the shot noise across the QPC, may give direct evidence for the trapping of fractional quasiparticles in the saddle-point quanutm dot.  

In conclusion, we have shown that QPCs in the quantum Hall regime can host Coulomb blockade physics even in the absence of an applied confining potential or disorder effects, resulting from the spontaneous formation of a quantum dot due to the interplay between the applied gate potential and interparticle  Coulomb interactions.  The discrepancy between the applied electrostatic potential and the total potential at the QPC provides direct evidence for edge state reconstruction.  By characterizing edge-state reconstruction in comparatively simple direct current transport measurements, moreover, we show that these effects may be identified and mitigated in experiments to probe fractional statistics in graphene based devices using QPCs\cite{zimmermann_tunable_2017,
cohen_nanoscale_2023,cohen_universal_2023} and Fabry-Perot interferometers\cite{ronen_aharonov-bohm_2021,deprez_tunable_2021,fu_aharonov-bohm_2023,werkmeister_strongly_2023,zhao_graphene-based_2022}. 

\begin{acknowledgments}
The authors thank J. Folk and A. Potts for comments on the manuscript.  
Work at UCSB was primarily supported by the Air Force Office of Scientific Research under award FA9550-20-1-0208 and by the Gordon and Betty Moore Foundation EPIQS program under award GBMF9471. 
LC and NS received additional support from the Army Research Office under award W911NF20-1-0082. 
TW and MZ were supported by the Director, Office of Science, Office of Basic Energy Sciences, Materials Sciences and Engineering Division of the U.S. Department of Energy under contract no. DE-AC02-05-CH11231 (van der Waals heterostructures program, KCWF16).
KK was supported by the U.S. Department of Energy, Office of Science, National Quantum Information Science Research Centers, Quantum Science Center. 
CR was supported by the National Science Foundation through Enabling Quantum Leap: Convergent Accelerated Discovery Foundries for Quantum Materials Science, Engineering and Information (Q-AMASE-i) award number DMR-1906325. 
K.W. and T.T. acknowledge support from JSPS KAKENHI (Grant Numbers 19H05790, 20H00354 and 21H05233).
\end{acknowledgments}

\normalem
\section{References}
\bibliographystyle{custom}
\bibliography{references}

\newpage
\clearpage

\renewcommand\thefigure{S\arabic{figure}}
\setcounter{figure}{0}

\end{document}


\title{Supplementary Information: Spontaneous localization at a potential saddle point from edge state reconstruction in a quantum Hall point contact}

\author{Liam A. Cohen}
\thanks{These authors contributed equally to this work}
\affiliation{Department of Physics, University of California at Santa Barbara, Santa Barbara CA 93106, USA}
\author{Noah L. Samuelson}
\thanks{These authors contributed equally to this work}
\affiliation{Department of Physics, University of California at Santa Barbara, Santa Barbara CA 93106, USA}

\author{Taige Wang}
\affiliation{Department of Physics, University of California, Berkeley, California 94720, USA}
\affiliation{Material Science Division, Lawrence Berkeley National Laboratory, Berkeley, California 94720, USA}
\author{Kai Klocke}
\affiliation{Department of Physics, University of California, Berkeley, California 94720, USA}
\author{Cian C. Reeves}
\affiliation{Department of Physics, University of California at Santa Barbara, Santa Barbara CA 93106, USA}
\author{Takashi Taniguchi}
\affiliation{International Center for Materials Nanoarchitectonics,
National Institute for Materials Science,  1-1 Namiki, Tsukuba 305-0044, Japan}
\author{Kenji Watanabe}
\affiliation{Research Center for Functional Materials,
National Institute for Materials Science, 1-1 Namiki, Tsukuba 305-0044, Japan}
\author{Sagar Vijay}
\affiliation{Department of Physics, University of California at Santa Barbara, Santa Barbara CA 93106, USA}
\author{Michael P. Zaletel}
\affiliation{Department of Physics, University of California, Berkeley, California 94720, USA}
\affiliation{Material Science Division, Lawrence Berkeley National Laboratory, Berkeley, California 94720, USA}
\author{Andrea F. Young}
\email{andrea@physics.ucsb.edu}
\affiliation{Department of Physics, University of California at Santa Barbara, Santa Barbara CA 93106, USA}

\date{\today}
\maketitle
\onecolumngrid

\section{Methods}

\subsection{Measurement}
Experiments were performed in a dry dilution refrigerator with a base temperature of ~20mK.  Electronic filters are used in line with transport and gate contacts in order to lower the effective electron temperature.  To improve edge mode equilibration to the contacts most measurements are performed at 300mK unless otherwise noted.  Electronic measurements were performed using standard lock-in amplifier techniques.  For the diagonal conductance measurements an AC voltage bias at 17.77Hz is applied via a 1000x resistor divider to (see supplementary Fig.~1 for contact references) C3-4 and the resulting current is measured using an Ithaco 1211 trans-impedance amplifier on C5-6 with a gain of $10^{-7}$ A/V.  The voltage is measured between contacts C1-2 and C7-8 with an SR560 voltage pre-amplifier with a gain of 1000.  For two terminal measurements the same AC bias is applied to contacts C1-4 and the current is measured via C5-8.  DC bias was added on top of the AC bias using a passive summer.

We use transport contacts in pairs to decrease the contact resistance - improving transport quality. The diagonal conductance is determined by exciting an AC voltage on contacts C3/C4, and then measuring the resulting AC current $I_\mathrm{out}$ on contacts C5/C6. The diagonal voltage drop, $V_D$, is measured from C1/C2 to C7/C8. The differential diagonal conductance is then defined as $G_D = I_\mathrm{out}/V_D$.

\begin{figure*}[ht]
    \centering
    \includegraphics[width=100mm]{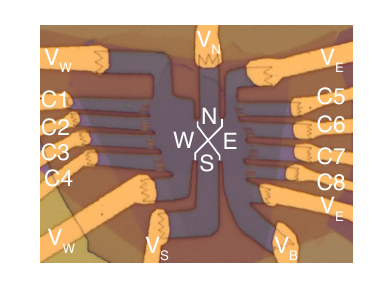}
    \caption{\textbf{Optical Micrograph of Measured Device} Transport contacts to the monolayer are  labeled C1-8, while the gates are labeled by their corresponding control voltages $V_i$.}
    \label{fig:supp_device}
\end{figure*}

\subsection{Thomas-Fermi Simulation}

We consider a classical effective model wherein the electron density $n(\mathbf{r})$ of the two-dimensional electron gas adjusts according to the local electrostatic potential and compressibility of an interacting Landau-level (LL).
The classical energy functional can be decomposed into the Hartree energy, the interaction with an externally applied  potential $\Phi(\mathbf{r})$, and the remainder,
\be
    \begin{aligned}
        \mathcal{E}[n(\mathbf{r})] &= \frac{e^2}{2}\int_{{\bf{r_1}, \bf{r_2}}} n({\bf{r_1}}) V({\bf{r_1}, \bf{r_2}}) n({\bf{r_2}})\\
        & \qquad \qquad - e\int_{\mathbf{r}} \Phi(\mathbf{r}) n(\mathbf{r}) + \mathcal{E}_{xc}[n(\mathbf{r})],
    \end{aligned}
\ee

$\mathcal{E}_{xc}[n]$ contains not only the exchange-correlation energy but also single-particle contributions (e.g. inter-LL interactions, disorder, Zeeman energies, etc.). If $n(\mathbf{r})$ varies slowly compared with $\ell_{B}$, we may neglect the dependence of the functional $\mathcal{E}_{xc}[n(\mathbf{r})]$ on the gradient $\boldsymbol{\nabla} n$ and employ the local density approximation (LDA),
\begin{equation}
    \mathcal{E}_{xc}[n(\mathbf{r})] = \int_{\mathbf{r}} E_{xc}(n(\mathbf{r}))
\end{equation}
where $E_{xc}(n)$ is determined for a system at \textit{constant} density $n$.
Our aim here is to find the density configuration $n(\mathbf{r})$ corresponding to the global minimum of the free energy $\mathcal{E}[n(\mathbf{r})]$.

The geometry we consider is shown in supplementary Fig.~3a-b.
There are four top gates a distance $d_t$ above the sample and one back gate a distance $d_b$ below, between which the space is filled by hBN with dielectric constant $\epsilon_{\perp} = 3$ and $\epsilon_{\|} = 6.6$ \cite{laturia_dielectric_2018}.
The N/S gates and the E/W gates are shifted by $w_{NS}$ and $w_{EW}$, respectively, from the center (see supplementary Fig.~3a-b).
We make an approximation by treating the cut-out ``X''-shaped region to define the gates as  a metal held at fixed voltage $V=0$ (rather than as a vacuum). 
This allows us to analytically solve for the electrostatic Green's function and gate-induced potentials without resorting to e.g. COMSOL simulations.

For a coarse-grained system with a finite resolution grid, the classical energy functional
becomes
\begin{equation}
    \begin{aligned}
	    \mathcal{E}[\{n(\mathbf{r})\}] &= \mathcal{E}_C + \mathcal{E}_{xc} + \mathcal{E}_{\Phi}\\
	    \mathcal{E}_C &= \frac{1}{2 A} \sum_{\mathbf{q}} V(\mathbf{q}) n(\mathbf{q}) n(-\mathbf{q}) \\
	    \mathcal{E}_{xc} &= \sum_{\mathbf{r}} E_{xc}(n(\mathbf{r})) \dif A,\\
	    \mathcal{E}_{\Phi} &= \sum_{\mathbf{r}} \Phi(\mathbf{r}) n(\mathbf{r}) \dif A
	\end{aligned}
	\label{eq:energy_functional}
\end{equation}
where $\dif A = \dif x \dif y$ is the grid area, $A$ is the total area, and $n(\mathbf{q}) = \sum_{\mathbf{r}} e^{-i \mathbf{q} \cdot \mathbf{r}} n(\mathbf{r}) \dif A$.
$\mathcal{E}_C$ is set by the gate-screened Coulomb interaction,
\begin{equation}
    V(\mathbf{q})= \frac{e^2}{4 \pi \epsilon_0 \epsilon_{\mathrm{hBN}}} \frac{4 \pi \sinh \left( \beta d_t|\mathbf{q}| \right) \sinh \left(\beta d_b|\mathbf{q}|\right)}{\sinh \left(\beta(d_t + d_b)|\mathbf{q}| \right) |\mathbf{q}|},
    \label{eq:gate_potential}
\end{equation}
$\epsilon_{\mathrm{hBN}} = \sqrt{\epsilon_{\perp} \epsilon_{\|}}$ and $\beta = \sqrt{\epsilon_{\|} / \epsilon_{\perp}}$. 
$\mathcal{E}_{\Phi}$ is the one body potential term arising from the potential $\Phi(\mathbf{r})$ on the sample due to the adjacent gates,
\begin{equation}
    \Phi(\mathbf{q}) = - e V_{t}(\mathbf{q}) \frac{\sinh(\beta d_b|\mathbf{q}|)}{\sinh(\beta(d_t + d_b)|\mathbf{q}|)} - e V_B \frac{d_t}{d_t + d_b} 
\end{equation}
where $V_{t}(\mathbf{q})$ is the top gate potential and $V_B$ is the back gate potential.

The remaining energy $E_{xc}$ is defined by integrating the chemical potential $\mu$,
\begin{equation}
    E_{xc}(n) = \int_0^{n} \dif n' \mu(n').
\end{equation}
which  encodes information about the IQH and FQH gaps and the electron compressibility. For $\mu$ we use the experimentally  measured value obtained for monolayer graphene in the FQH regime ($B = 18 \, \si{T}$) we reported in an earlier work, Ref \cite{yang_experimental_2021}.

Physically, $n(\mathbf{r})$ can only vary on the scale of the magnetic length $\ell_B$ due to the underlying quantum Hall wavefunction. 
To capture this feature, we implement a square grid with periodic boundary conditions and meshing much finer than the scale of $\ell_B$,
and then evaluate $\mathcal{E}$ with respect to the Gaussian convoluted density profile \footnote{We first use an unbounded range for $n(\mathbf{r})$ to determine the external potential that fix bulk filling $\nu_{EW} = 1$ and $\nu_{NS} = 0$. Then we constrain $n(\mathbf{r})$ to fall in between $\nu = 0$ and $\nu = 1$ such that any transition from $\tilde{\nu} = 0$ to $\tilde{\nu} = 1$ in $\tilde \nu(\mathbf{r})$ has a finite width of scale $\ell_B$.}
\begin{equation}
    \tilde n(\mathbf{r}) = \mathcal{N}^{-1}\sum_{\mathbf{r}'} n(\mathbf r') e^{-\abs{\mathbf{r} - \mathbf{r}'}^2 / 2\ell_B^2},
\end{equation}
where $\mathcal{N}$ is the corresponding normalization factor.
We use a basin-hopping global optimizer with local L-BFGS-B minimization to vary $\{n(\mathbf{r})\}$ and find the lowest energy configuration.

Within this framework, we tune the E/W gate potentials such that $\nu_{EW}=1$ deep in the bulk and the N/S gate potentials such that $\nu_{NS}=0$. 
Various effects of reconstruction can then be explored by appropriately tuning the smoothness of the potential at the $\nu=0$ to $\nu=1$ interface (e.g. by tuning the channel width $w$, gate distances $d_{t,b}$ or gate voltages $\Phi_{g}$).

\section{Extended Information on Interaction-Driven Quantum Dots}
\begin{figure*}[ht]
    \centering
    \includegraphics[width = 170mm]{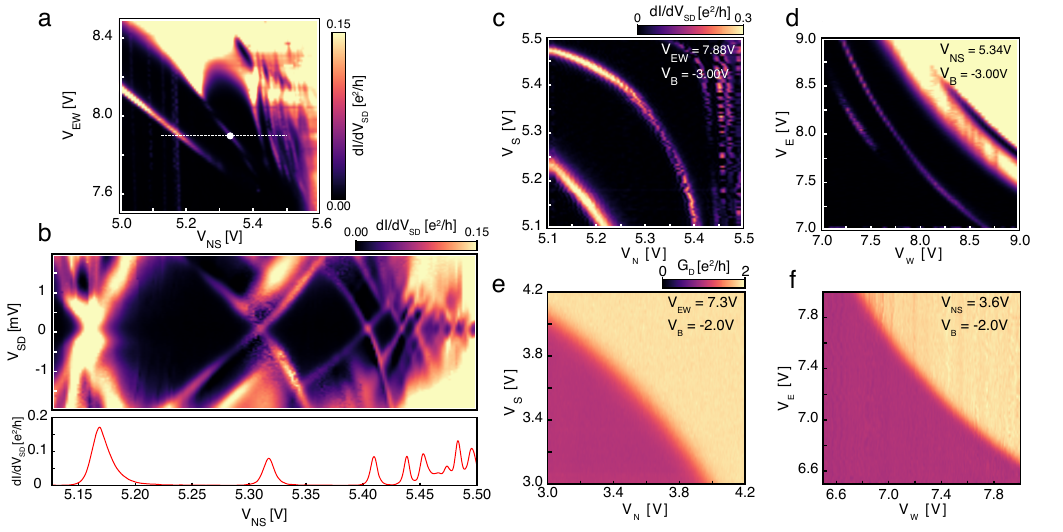}
    \caption{\textbf{Coulomb blockaded resonances on the electron side.} All data in this figure was taken at B = 9T and T = 20mK.  \textbf{(a)} Two terminal conductance across the device plotted against $V_{NS}$ and $V_{EW}$.  \textbf{(b)} Differential conductance versus source-drain bias plotted along the white dashed line in (a) as well as the corresponding zero-bias line cut.  Along the white dashed line $\nu_{EW} = 3$ and $\nu_{NS} = 0$. \textbf{(c-d)} Two terminal conductance plotted against $V_{N(W)}$ and $V_{S(E)}$ for the Coulomb blockaded resonances marked by the white dashed line in (a). The primary resonance shown in (d) is demarcated by the white dot in (a). \textbf{(e-f)} Diagonal conductance across the device plotted against $V_{N(W)}$ and $V_{S(E)}$ for the transmission step between $G_D = 1$, and $G_D = 2$. } 
    \label{fig:supp_saddle_point_dot}
\end{figure*}

The Coulomb blockaded resonant structure is fairly ubiquitous in this device. Fig.~S\ref{fig:supp_saddle_point_dot} presents a series of several resonances which exhibit Coulomb blockade on the electron side of the device at 9T and 20mK.  Fig.~S\ref{fig:supp_saddle_point_dot}a shows the resonances in the ($V_{NS}$, $V_{EW}$) plane, where $\nu_{EW} \sim 3$ and $\nu_{NS} \sim 0$.  The chemical potential of the quantum dot is modulated directly with the N/S gates as $V_{NS}$ is swept along the dashed white line in Fig.~S\ref{fig:supp_saddle_point_dot}a.  Fig.~S\ref{fig:supp_saddle_point_dot}b shows the differential conductance as a function of $V_{NS}$ and $V_{SD}$ along this $V_{NS}$ trajectory along with the zero-bias cut.  

The sharp peaks in Fig.~S\ref{fig:supp_saddle_point_dot}b have some finite curvature in the ($V_{SD}$, $V_{NS}$) plane. Since the slope here is a direct measure of $-C_{NS} / C_{\Sigma}$, where $C_{\Sigma}$ is the total capacitance of the dot, this indicates that as $V_{NS}$ is increased, the capacitance of the N/S gates to the dot is decreasing relative to $C_{\Sigma}$.  This is consistent with the dot being squeezed in the N/S direction as $V_{NS}$ is increased.  This can be further corroborated by the observation that as $V_{NS}$ is increased in Fig.~S\ref{fig:supp_saddle_point_dot}a, the slope of successive resonances decreases, indicating a decreased sensitivity to modulations in $V_{NS}$ -- i.e., $C_{EW} > C_{NS}$ as $V_{NS}$ increases.  Additionally, Fig.~S\ref{fig:supp_saddle_point_dot}c-d shows a representative resonance as as function of $V_N$ vs. $V_S$ as well as $V_E$ vs. $V_W$.  Much like the analysis in the main text (Fig.~2), when $V_N = V_S$ or $V_E = V_W$, the slope in the $V_N$/$V_S$ plane or the ($V_E$, $V_W$) plane is near $-1$, indicating the dot is roughly centered in the QPC.

The behavior of a monotonic transition in $G_D$ as a function of $V_{N}$ vs. $V_{S}$ and $V_{E}$ vs. $V_{W}$ can be seen in Fig.~S\ref{fig:supp_saddle_point_dot}e-f.  It has been well established that monotonic steps in conductance at a quantum Hall QPC can be described by the scattering of electrons in a magnetic field at a saddle point potential \cite{floser_transmission_2010}.  Qualitatively, the behavior of the monotonic transmission step, and the Coulomb blockaded resonances in \ref{fig:supp_saddle_point_dot}c-d are similar; the monotonic transmission step and the resonance both have the same curvature in the ($V_N$, $V_S$) or ($V_E$, $V_W$) plane.  This further verifies the conclusion that the quantum dots are sitting on top of an externally applied potential which forms a saddle point, not a 2D well.

\section{Thomas-Fermi calculation of the local density in the QPC}
\label{app:TF_cal}

\begin{figure*}[ht]
    \centering
    \includegraphics[width=0.95\textwidth]{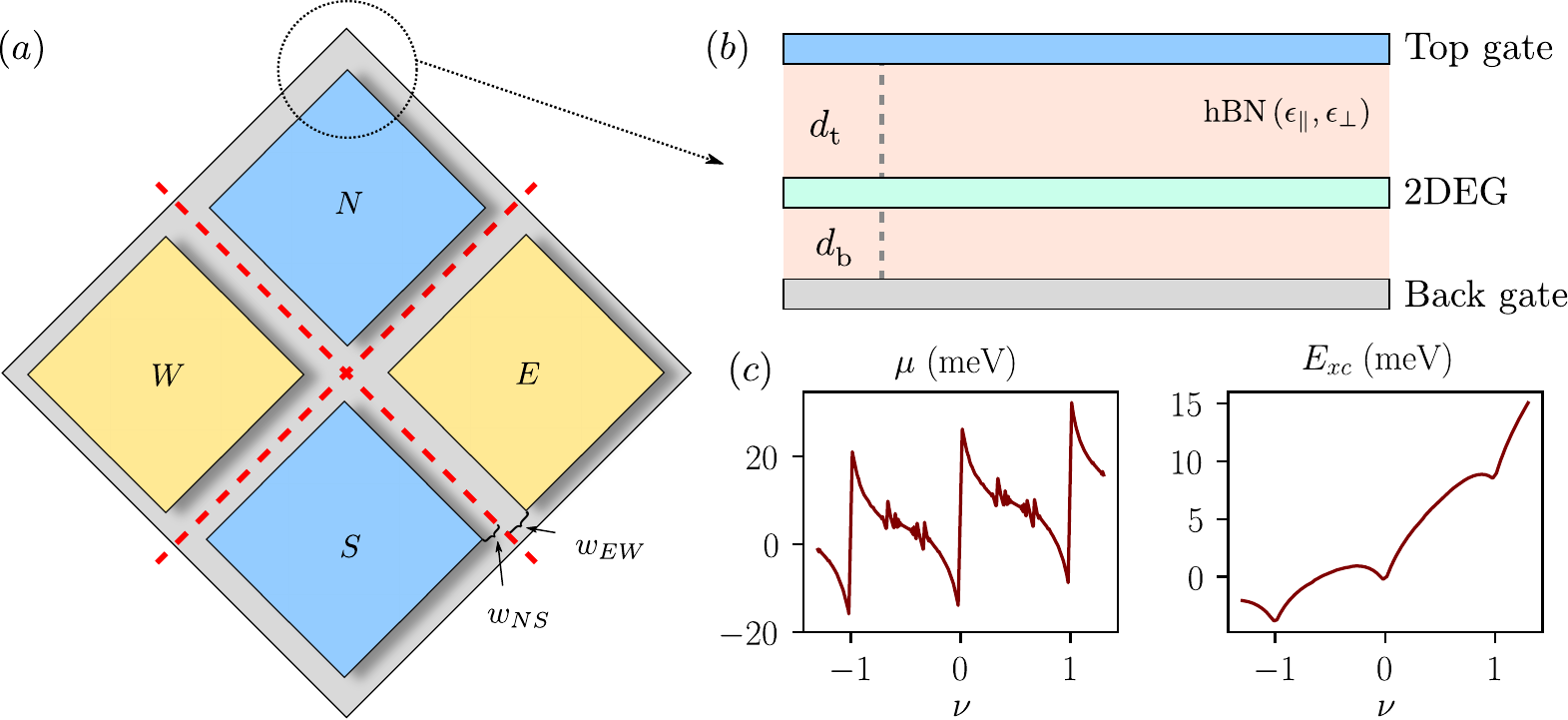}
    \caption{
    \textbf{Effective model considered within the Thomas-Fermi framework.}
    \textbf{(a)} Top-view of the gate setup, which involves four square gates labeled N,S,E, and W respectively, each of which is held at a potential denoted by $V_{N,S,E,W}$.
    The back-gate (gray) is held at constant $V_B$.
    We allow for the gate displacement from the center to be different for E/W and N/S gates.
    \textbf{(b)}
    Side-view of the gates reveals two distances $d_t$ and $d_b$ for the separation between the 2DEG and the top gate or bottom gate, respectively.
    \textbf{(c)} 
    The chemical potential $\mu$ and internal energy $E_\textrm{xc}$ derived from Ref.~\cite{yang_experimental_2021} which is taken as input data for the Thomas-Fermi calculation carried out here. 
    }
    \label{fig:tf_model_summary}
\end{figure*}

\subsection{Fractional Reconstruction}

\begin{figure*}[ht!]
    \centering
    \includegraphics[width = 120mm]{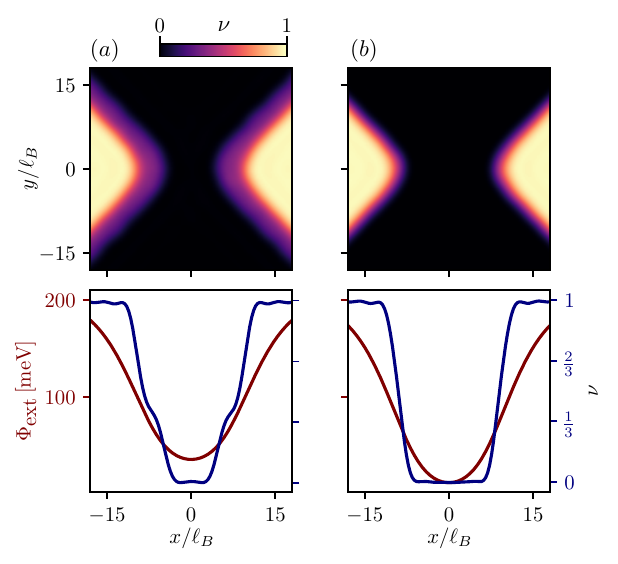}
    \caption{
    \textbf{Extended reconstructed electron density profiles in a QPC geometry in the fractional regime}. 
    (a-b) Fractional reconstruction in an integer bulk filling ($\nu=1$) with $E_C = \,46.4~\si{meV}$ in the regime where the confining potential is sharper than presented in main text Fig.~3a.
    \textbf{(a)} With $E_V / E_C = 0.50$, the $\nu=\frac13$ island disappears and the $\nu=\frac13$ strips become narrower.
    \textbf{(b)} With $E_V/E_C = 0.57$, the $\nu=\frac13$ strips disappear completely.
    }
    \label{fig:TF_fractional2}
\end{figure*}

To approximate the experimental device, we simulate a system with magnetic field $B = 13 \,\si{T}$ (magnetic length $\ell_B = 7\,\si{nm}$) \footnote{In the experiment \cite{yang_experimental_2021}, $\mu$ was obtained at $B = 18 \, \si{T}$, which introduces a difference in energy scale of $E_C$ by a factor of 1.19 relative to $B = 13 \, \si{T}$ at which we carry out our calculations}, gate distances $d_t = d_b = 30 \, \si{nm}$ and channel widths $w_{EW} = w_{NS} = 35\, \si{nm}$ on a $40 \ell_B \, \times \, 40 \ell_B$ system with gridsize $\dif x = \dif y = \tfrac14 \ell_B$ \footnote{It is advantageous to take a smaller $d_t$ here than the experimental $60 \,\si{nm}$ so that the necessary system size $L$, and hence the number of variational parameters in the optimization, remains tractable.}.

For a finite system and mesh size, we find that edge reconstruction is most easily observed when $E_{xc}$  is increased by a factor of $1 - 2$  (for the data shown here, $E_{xc}$ is increased by 1.8x).  One possible origin of this discrepancy is screening from the filled LLs. 
In Ref.~\cite{yang_experimental_2021} it was found the screening from virtual inter-LL  transitions (which is implicitly included in our $\mathcal{E}_{xc}$, since we use experimental data) reduces the IQH gaps, and hence features in $\mu$, by about a factor of 1.4 relative to their unscreened values.  However, our phenomenological model does \emph{not} account for inter-LL screening in the electrostatic contributions $\mathcal{E}_C$ and $\mathcal{E}_\Phi$. 
Decreasing their scale relative to $E_{xc}$ can thus be interpreted as a very crude implementation of LL-screening in the remaining energy functionals. Of course other approximations are in play as well,  e.g. the local density approximation, so at the outset we do not expect more than phenomenological agreement with experiment. 

At a translationally invariant boundary between $\nu=1$ to $\nu=0$, previous variational calculations have suggested edge-state reconstruction makes it energetically favorable to form an additional strip of $\nu=\tfrac13$ \cite{khanna_fractional_2021}.  However, in a QPC geometry where translation symmetry is broken, the effects of reconstruction are much richer and consequently require numerical treatment to fully characterize.  In Fig.~S6 of the main text, we demonstrate the effects of edge-reconstruction in a QPC as a function of the potential smoothness quantified by the ratio $E_V / E_C$.  To accomplish this, we fix $V_{NS} = -V_B$ and tune the smoothness of the confining potential via the relative magnitudes of $V_{EW}$ and $V_B$. 
For the smallest value of $E_V / E_C$, we find in main text Fig.~3a reconstructed $\nu=\frac13$ strips which extend far enough into the QPC that they merge, corresponding to a fractional conductance across the QPC ($V_{EW}=0.51\,\si{V}$, $V_B=-0.19\,\si{V}$). With increasing $E_V / E_C$, in main text Fig.~3b the $\nu=\tfrac13$ strips become narrower and an island of fractional filling is formed in the center of the QPC ($V_{EW}=0.60\,\si{V}$, $V_B=-0.10\,\si{V}$).  As shown in Fig.~S\ref{fig:TF_fractional2}a, we confirmed upon further increasing $E_V / E_C$ ($V_{EW} = 0.61\,\si{V}$, $V_B=-0.09\,\si{V}$), the $\nu=\tfrac13$ strips become even narrower and the island disappears. For extremely sharp edges ($V_{EW}=0.70\,\si{V}$, $V_B=0.0\,\si{V}$), in Fig.~S\ref{fig:TF_fractional2}b the $\nu=\tfrac13$ strips disappear entirely leaving the expected density profile for a $\nu = 1$ edge everywhere in the QPC.

We note that the $\nu=\tfrac13$ strips are smoothly connected to the bulk (i.e. not separated by a region of $\nu=0$) even for very smooth confining potentials, which is consistent with variational calculations which more accurately account for the energetics of $\nu=1$ to $\tfrac13$ edges beyond the local-density approximation \cite{khanna_fractional_2021}. 
For a line cut taken directly across the $\nu=1$ to $\nu=0$ interface (e.g. between the east and north gates), the confining potential $\Phi_\textrm{ext}$ gives a confinement energy $E_V \approx 0.44 E_C$ and $0.49 E_C$ for the two scenarios in main text Fig.~3a-b respectively, and $0.50 E_C$ for the scenario where the island of fractional filling disappears (Fig.~S\ref{fig:TF_fractional2}a).
The gate potentials reported here are slightly smaller than actual values used in the experiment. 
In the experiment, the bulk filling far from the QPC can be as high as $\nu=\pm 4$ even though the local filling at the QPC will give $\nu=1$ to $\nu=0$ interfaces as we simulate here (see Fig.~1 of the main text).
Since we simulate only the small region close to the QPC, the effective gate voltages we employ will necessarily reflect a smaller range than that of the experiment.

In generic quantum Hall systems, it is widely believed that disorder and localized impurities are responsible for the formation of localized states in the QPC which facilitate resonant tunneling and are responsible for the non-monotinicity observed in conductance measurements across QPC transmission steps \cite{Milliken1996,Ando1998,Baer2014}. 
In the device studied in this paper, we do not expect such disorder, and the localized states were attributed to an entirely intrinsic mechanism based purely on including the Coulomb interaction at the soft gate-defined edges of the QPC.  That picture is supported by our Thomas-Fermi calculations, where we have shown explicitly that the Coulomb interaction itself is sufficient to favor reconstruction of a local island.  We see then that by appropriate tuning of the gate voltages, one can reproduce a Coulomb blockade and fractional transmission in close approximation to the experimental findings of this report.

\subsection{Integer Reconstruction}
\begin{figure*}[ht!]
    \centering
    \includegraphics[width = 170mm]{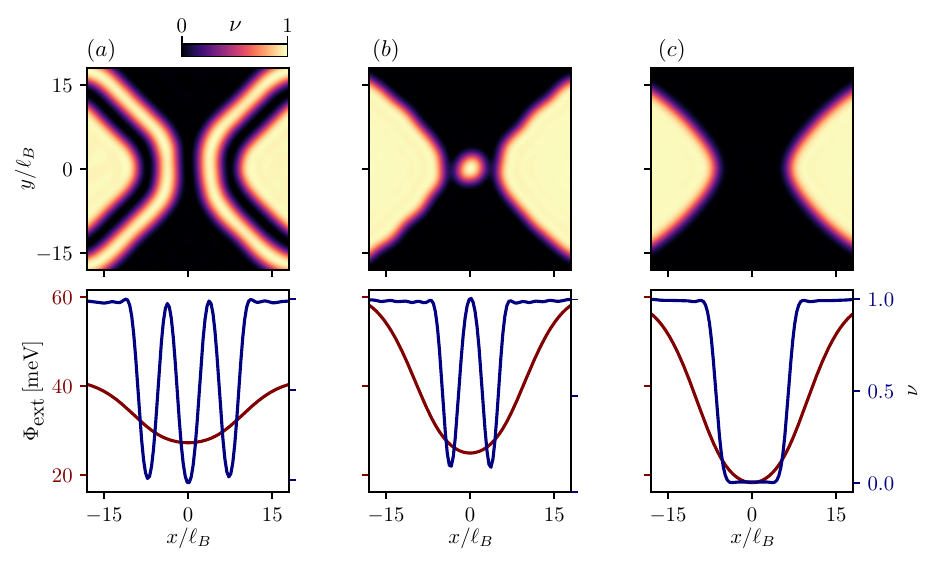}
    \caption{
    \textbf{Reconstructed electron density profiles in a QPC geometry in the integer reconstruction regime}. 
    (a-c) Integer reconstruction in an integer bulk filling ($\nu=1$) with $E_C = \,11.6~\si{meV}$ as the confining potential smoothness is varied.
    \textbf{(a)} With $E_V / E_C = 0.18$, the reconstructed $\nu=1$ stripes extend along the entire edge.
    \textbf{(b)} With $E_V/E_C = 0.47$, the reconstructed $\nu=1$ stripes shrink to a single circular dot with $\nu=1$ in the center of the QPC.
    \textbf{(c)} With $E_V/E_C = 0.55$, no reconstruction is observed.
    }
    \label{fig:TF_Integer}
\end{figure*}

Now we turn our attention to the possibility of reconstructed strips and islands with integer filling.  Integer reconstruction can be made more energetically favorable in the numerics by reducing the magnitude of $E_C$ such that redistributing integer charge along the edge out competes the energy gained from from forming a correlated FQH strip. Fixing $\ell_B$, $d_{t,b}$ and $w_{EW, NS}$ as before, we now take $E_C = \,11.6~\si{meV}$, and work on a system grid which is $40\ell_B \times 40 \ell_B$ large.

In Fig.~S\ref{fig:TF_Integer}, we demonstrate edge-reconstruction as a function of the confining potential smoothness tuned via the gate voltages. For the smallest value of $E_V / E_C$, we find in Fig.~S\ref{fig:TF_Integer}a reconstructed $\nu=1$ strips that exist throughout the QPC,
($V_{EW}=43.8\,\si{mV}$, $V_{NS}=107\,\si{mV}$, $V_B=107\,\si{mV}$).  With increasing $E_V / E_C$, in Fig.~S\ref{fig:TF_Integer}b the formation of an island of integer filling within the QPC ($V_{EW} = -143\,\si{mV}$, $V_{NS}=79.6\,\si{mV}$, $V_B=79.6\,\si{mV}$) is observed, and in Fig.\ref{fig:TF_Integer}c, for the largest value of $E_V/E_C$, a monotonic transition from $\nu = 0$ to $\nu = 1$ is recovered everywhere in the QPC ($V_{EW} = -167\,\si{mV}$, $V_{NS}=55.7\,\si{mV}$, $V_B=55.7\,\si{mV}$).

In contrast to the $\nu=\frac13$ strips, the reconstructed $\nu=1$ strips in Fig.~S\ref{fig:TF_Integer}a are separated from the bulk by fully depleted regions, which is also consistent with variational calculations \cite{khanna_fractional_2021}. For a line cut taken directly across the $\nu=1$ to $\nu=0$ interface (e.g. between the east and north gates), the confining potential $\Phi_\textrm{ext}$ gives a confinement energy $E_V \approx 0.18 E_C, 0.47 E_C$ and $0.55 E_C$ for three scenarios Fig.~S\ref{fig:TF_Integer}a-c respectively. We leave a more detailed study of the interplay of integer and fractional reconstruction in this QPC geometry to future work.

\section{Additional Experimental Evidence for Fractional Reconstruction at the QPC}
\begin{figure*}[ht]
    \centering
    \includegraphics[width = 170mm]{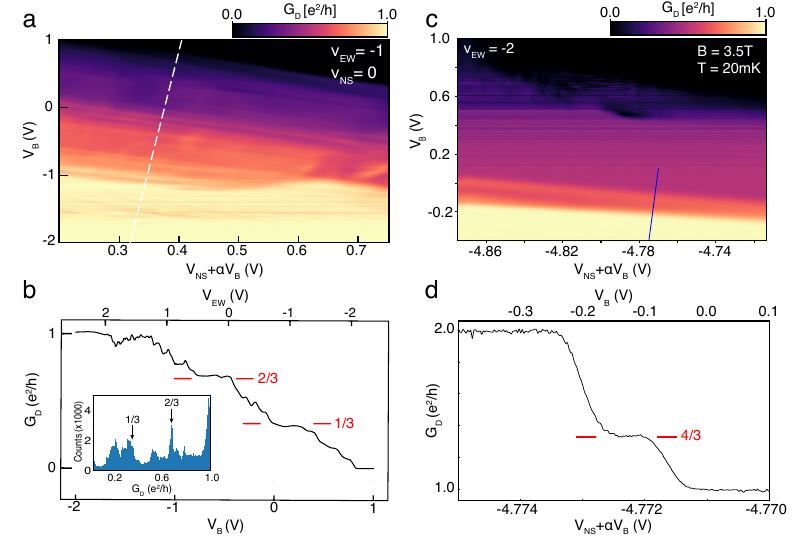}
    \caption{\textbf{Evidence for fractional edge modes at the boundary of a bulk integer state:} \textbf{(a)} $G_D$ at a fixed $\nu_{EW} = -1$ versus $V_{NS} + \alpha V_B$ and $V_{B}$.  Here $V_{NS}$ is swept in a range dependent on $V_{B}$ such that the range $V_{NS} + \alpha V_B$ is kept fixed.  Additionally, as $V_B$ is swept, $V_{EW}$ is swept concomitantly to keep the bulk filling factor, set by $V_{EW} + \alpha V_B$, fixed at $\nu = -1$, but varying $V_{EW} - V_B$. 
    The x-axis range corresponds to the full width of the $\nu=0$ plateau.  \textbf{(b)} The diagonal conductance map of the $\nu_{EW} = -1$ state at B=13T presented in panel (a) shows a number of features between $G_D = 1$ and $G_D = 0$. The statistical frequency of conductance values across the entire plot, reveals a number of sharp peaks corresponding to conductance plateaus between 0 and 1. The most prominent occurs at $G_D = 2/3$. A broader peak is observed near 1/3, though the quantization is less exact, and several further sharp peaks are seen centered near 0.2, 0.5, and 0.8. The origin of these peaks is unknown, and may be due to a more complicated reconstructed edge structure than is accounted for by any model discussed in this work. \textbf{(c)} Diagonal conductance map with the E/W regions in a fixed filling factor $\nu_{EW}=-2$ and N/S regions entirely within the $\nu_{NS}=0$ plateau. On the y-axis, the bottom gate and E/W gates are swept in opposite directions to vary $(V_B-V_{EW})$ while keeping $\nu_{EW} = -2$ fixed. On the x-axis, the N/S gates are swept across the entire $\nu=0$ plateau. \textbf{(d)} Line cut given in (c) showing two integer plateaus and an intermediate fractional plateau near $G_D = 1.33$ demarcated by the red lines.  In both (a) and (b) a correcting scale factor of $0.95$ is applied uniformly to $G_D$ to correct for a parallel conduction channel created in our device due to fringe doping from the Si gate, leading to over-quantized plateaus.  Here only one $e/3$ edge mode is observed, in contrast to the data at B = 13T where two $e/3$ modes are observed.  This is similar to previous reports in GaAs \cite{bhattacharyya_melting_2019}, however, the reason why one plateau is favored at low field compared to $B = 13T$ remains unclear.}
    \label{fig:fqh_reconstruction}
\end{figure*}
In Fig.~\ref{fig:fqh_reconstruction} we present data demonstrating additional plateaus in transmission across the qpc while the bulk filling factor is kept in $\nu = 1$.  The presence of these plateaus provide strong evidence for the existence of an additional incompressible strip at fractional filling within the QPC at low values of $E_V / E_C$ consistent with our simulations.

\section{Experimental Realization of the Unreconstructed Limit of $\nu = 1$ Transmission at the QPC}
\begin{figure*}[ht]
    \includegraphics[width = 170mm]{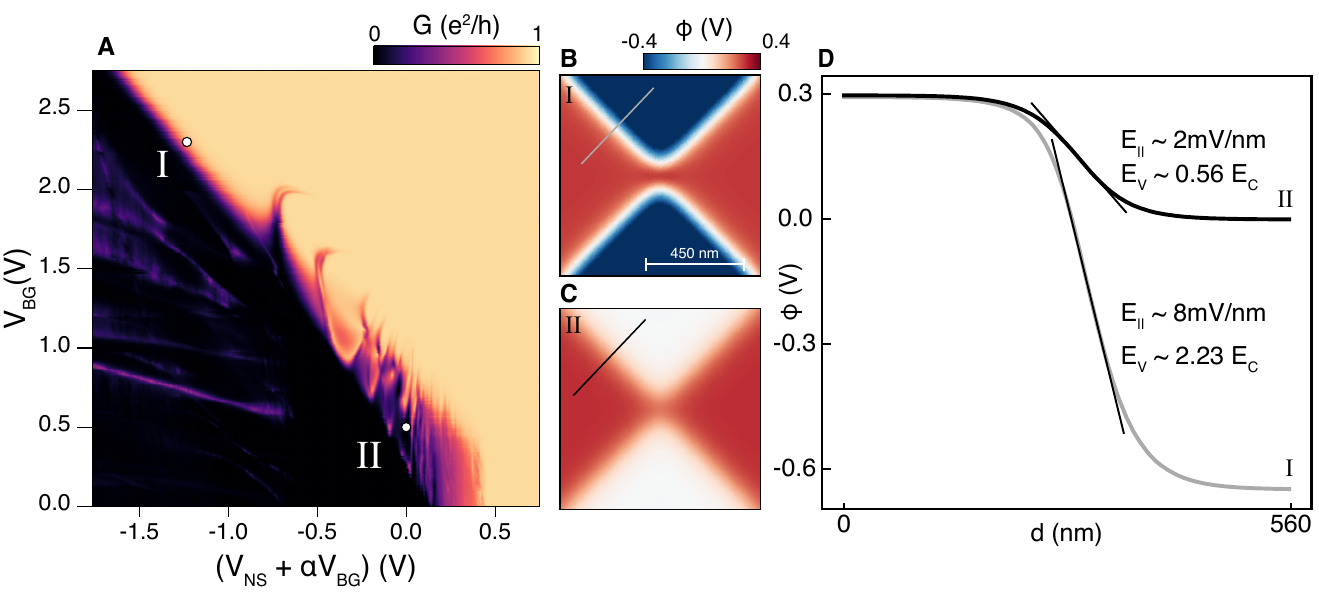}
    \caption{
    \textbf{Signatures of reconstruction in $G$ between $\nu = 1$ edge modes.} 
    \textbf{(a)} The conductance measured across the QPC with both the East and West regions in $\nu = 1$ at B = 8T. The East and West gate voltages are swept in the opposition direction of $V_{\text{BG}}$ along the y axis, through the range $V_\mathrm{EW} \in (0.6V, -2.191V)$, to maintain a fixed filling factor while varying the voltage difference and thereby the potential sharpness.
    \textbf{(b)} The simulated electric potential at the monolayer, corresponding to the operating point I.
    \textbf{(c)} Same as (b) but at the operating point II, where the potential is much softer.
    \textbf{(d)} Simulated potential along the contours marked in grey and black in panels B and C, respectively. The softness is quantified by the maximum magnitude of the in-plane confining electric field, $E_{\parallel}$ (\textit{i.e.} simply the gradient of the potential normal to the boundary between the N(/S) and E(/W) regions).
    \label{fig:electrostatics_integer}
    }
\end{figure*}
In Fig.~\ref{fig:electrostatics_integer} we present data on a second device measured in \cite{cohen_universal_2023} which looks at the transmission as a function of potential sharpness for the $\nu = 1$ state.  Due to higher gate limits in this device, by applying a large voltage difference between the back-gate and the E/W gates, we were able to achieve the $E_V / E_C > 1$ limit.  In this limit the presence of spontaneously formed quantum dots from edge reconstruction begins to vanish and a single monotonic step from full pinch off to $G = e^2/h$ is recovered.

\normalem
\section{References}
\bibliographystyle{custom}
\bibliography{references, recon_bib}